\pdfoutput=1
\documentclass[amsmath,amssymb,aps,twocolumn,superscriptaddress]{revtex4-2}
\usepackage{amsmath}
\usepackage{amssymb}
\usepackage{bm}
\usepackage{graphicx}
\usepackage{epsf}
\usepackage{xcolor}
\usepackage{color}
\usepackage{graphicx}
\usepackage[version=4]{mhchem}
\usepackage[bookmarks=false]{hyperref}
\usepackage{siunitx}
\usepackage[english]{babel}
\usepackage[T1]{fontenc}
\usepackage{dcolumn}
\usepackage{tabularx}
\usepackage{array} 
\usepackage{booktabs}

\makeatletter
\def\maketitle{
\@author@finish
\title@column\titleblock@produce
\suppressfloats[t]}
\makeatother
\newcommand{\beginsupplement}{
    \setcounter{table}{0}
    \renewcommand{\thetable}{S\arabic{table}}
    \setcounter{figure}{0}
    \renewcommand{\thefigure}{S\arabic{figure}}
    \setcounter{equation}{0}
    \setcounter{section}{0}
    \renewcommand{\theequation}{S\arabic{equation}}
}

\begin{document}
\title{Strong coupling of polaritons at room temperature in a GaAs/AlGaAs structure}

\author{Hassan Alnatah}
\thanks{Address correspondence to: haa108@pitt.edu}
\affiliation{Department of Physics, University of Pittsburgh, 3941 O’Hara Street, Pittsburgh, Pennsylvania 15218, USA}

\author{Shuang Liang}
\altaffiliation{S.L. and H.A. contributed equally to this work}

\author{Qiaochu Wan}
\affiliation{Department of Physics, University of Pittsburgh, 3941 O’Hara Street, Pittsburgh, Pennsylvania 15218, USA}

\author{Jonathan Beaumariage}
\affiliation{Department of Physics, University of Pittsburgh, 3941 O’Hara Street, Pittsburgh, Pennsylvania 15218, USA}

\author{Ken West}
\affiliation{Department of Electrical Engineering, Princeton University, Princeton, New Jersey 08544, USA}

\author{Kirk Baldwin}
\affiliation{Department of Electrical Engineering, Princeton University, Princeton, New Jersey 08544, USA}

\author{Loren N. Pfeiffer}
\affiliation{Department of Electrical Engineering, Princeton University, Princeton, New Jersey 08544, USA}

\author{Man Chun Alan Tam}
\affiliation{Department of Electrical and Computer Engineering, University of Waterloo, Waterloo, ON, Canada}
\affiliation{Waterloo Institute for Nanotechnology, University of Waterloo, Waterloo, ON, Canada}

\author{Zbigniew R. Wasilewski}
\affiliation{Department of Electrical and Computer Engineering, University of Waterloo, Waterloo, ON, Canada}
\affiliation{Waterloo Institute for Nanotechnology, University of Waterloo, Waterloo, ON, Canada}
\affiliation{The Institute for Quantum Computing (IQC), University of Waterloo, Waterloo, ON, Canada}

\author{David W. Snoke}
\affiliation{Department of Physics, University of Pittsburgh, 3941 O’Hara Street, Pittsburgh, Pennsylvania 15218, USA}

\date{\today}

\begin{abstract}
We report direct measurement of the dispersion relation of polaritons in GaAs/AlGaAs microcavity structures at room temperature, which clearly shows that the polaritons are in the strong coupling limit. The Rabi splitting of the polariton states decreases as the polariton gas increases in density, but even when the polariton gas becomes a coherent, Bose-condensate-like state, the polaritons retain a strong exciton component, as seen in the nonlinear energy shift of the light emission. This opens up the possibility of polaritonic devices at room temperature in a material system which can be grown with very high quality and uniformity. 
\end{abstract}

\maketitle

\section{INTRODUCTION}
Polaritons can be viewed as photons dressed with an effective mass and repulsive interactions, due to the strong coupling of a cavity photon state and a semiconductor exciton state. Over the past two decades numerous experiments have demonstrated Bose condensation and nonlinear coherent effects in GaAs/AlGaAs structures at cryogenic temperatures (e.g. \cite{deng2002condensation,kasprzak2006bose,balili2007bose,abbarchi2013macroscopic,sanvitto2010persistent,lagoudakis2009observation}).  Strong coupling occurs when the Rabi splitting is larger than the half-widths of the states, 
implying that the vacuum oscillation rate between the exciton and photon modes is faster than their decay times.

Polaritonic strong coupling at room temperature is desirable to use their strong optical nonlinearity for practical polaritonic devices. Polaritons in the strong coupling limit have been seen in organic materials \cite{plumhof2014room,dusel2020room,wei2022optically} and perovskites \cite{su2020observation,su2018room}. Significant efforts have also been made to achieve strong coupling in III-V structures at room temperature. Earlier results \cite{tsintzos2009room,brodbeck2013room,suchomel2017room} had spectral half-widths comparable to, or only slightly smaller than, the measured Rabi splittings. Furthermore, none of these earlier works above showed nonlinear shift of the spectral lines, line narrowing, or lasing as the excitation density increased. While a nonlinear line shift is not a direct indicator of strong coupling, it indicates that the polaritons have a significant excitonic fraction, and is essential for nonlinear optical applications.

In a previous paper, we reported polaritonic coherent emission at room temperature in GaAs/AlGaAs microcavity structures \cite{alnatah2024bose}. A thermalized energy distribution was seen, and it was clear that polaritonic effects involving an exciton fraction were present, as seen in the strong energy shift of the emission line with pump power. However, at that time, it was not clear what exciton fraction could be attributed to the polaritons; it was argued in Ref.~\cite{alnatah2024bose} that the coupling of the photons to excitons was weak but nonzero, with exciton fraction of 10\% or less, based on a model of the photon-exciton coupling using the observed photoluminescence (PL) data. 

We have now performed similar experiments 
with multiple structures with different designs, with different designed $Q$-factors of the cavities and different numbers of quantum wells, and a wide range of pumping conditions. Crucially, by looking at a sample with lower Q-factor of the microcavity, we can easily observe three polariton branches arising from the heavy-hole exciton, light-hole exciton, and cavity photon mode, which allows us to fit the entire dispersion curves of the polaritons.

In this work, we demonstrate a Rabi splitting of approximately 13 meV at room temperature, which is more than twice the heavy-hole exciton half-width of 5.2 meV. We find that in all of the samples studied, the polaritons are in strong coupling at low pump power at room temperature, and remain in relatively strong coupling as the pump power is increased into the coherent emission regime. The Rabi splitting decreases as density increases, but never fully collapses to zero, as far as we can measure. 

This result is surprising, but is a result of the fact that our samples have very sharp exciton lines in the quantum wells even at room temperature. (The Supplemental Material shows the exciton emission from bare quantum wells grown using the same process but without a microcavity.)
When the quantum wells are embedded in a microcavity, we expect that there will be splitting of the excitons into multiple states due to well width fluctuations, but this does not imply broadening of the exciton-polariton lines. The line width of polariton states primarily arises from the homogeneous broadening of the exciton states, and is not strongly affected by the presence of multiple exciton states.

\begin{figure*}
\includegraphics[width=0.6\textwidth]{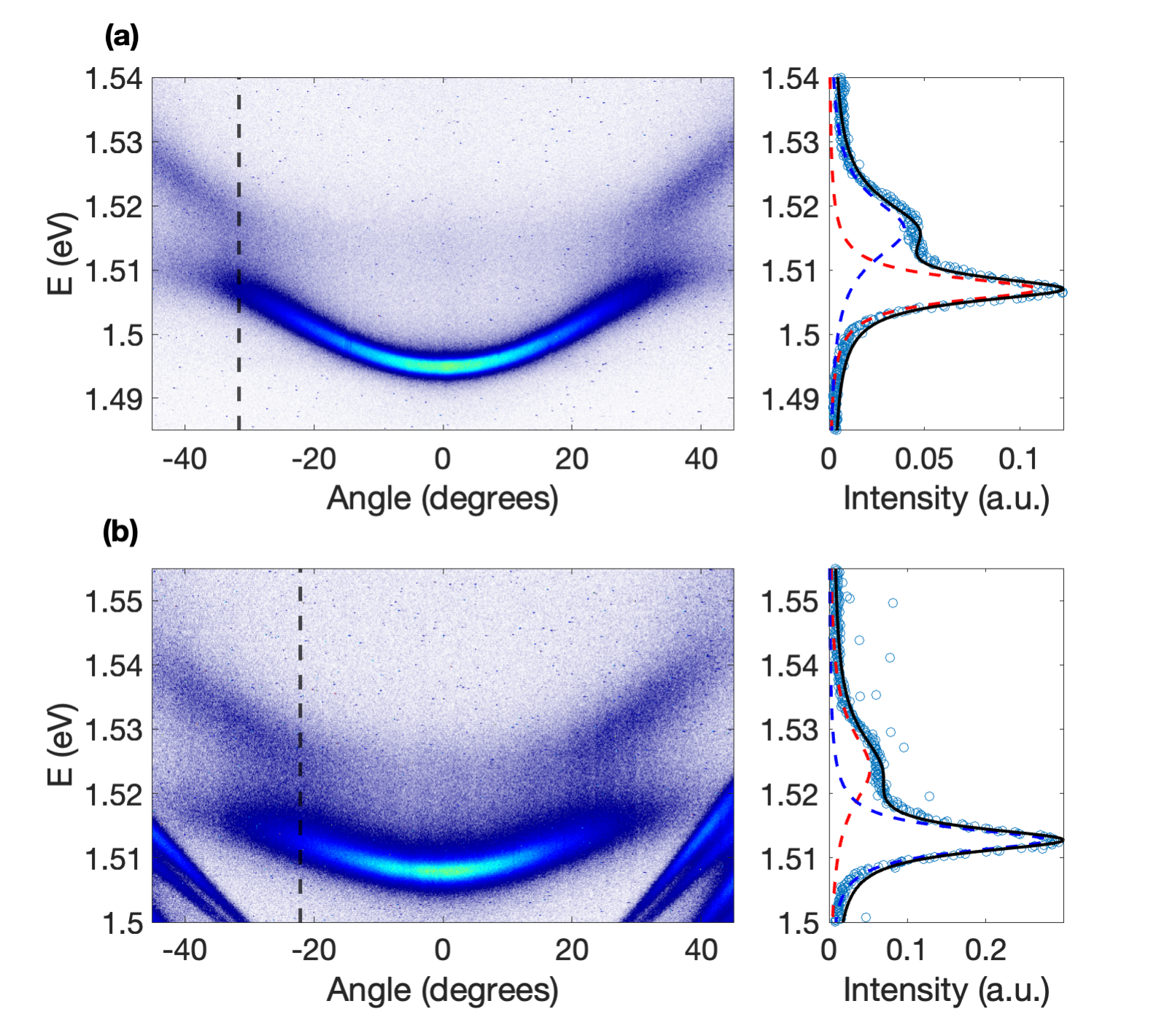}
\centering
\caption{\textbf{Strong coupling at room temperature} in Samples P1 and W1. {\bf (a)} Left plot: angle-resolved PL measurements of the polariton at very low pumping power for Sample P1 at a location with LP exciton fraction of approximately 0.12 at $k=0$. Right plot: the intensity as a function of energy along the slice at constant angle corresponding to the vertical dashed line in (a). To reduce the noise in the experimental data, we averaged over 20 pixels around the slice.  Dashed lines: fit to  two Lorentzian peaks. {\bf (b)} The same measurements at very low pumping power for Sample W1, for an LP exciton fraction of approximately 0.24 at $k=0$. The extra curved lines seen at low energy and high angle in (b) are ripples in the reflectivity spectrum at the edge of the stopband of the DBR cavity of this sample. }
\label{fig:strong-coupling}
\end{figure*}

\begin{table}[b]  
    \centering
    \renewcommand{\arraystretch}{1.5}
    \begin{tabularx}{0.45\textwidth} { 
    | >{\centering\arraybackslash}m{2.5cm} 
    | >{\centering\arraybackslash}X 
    | >{\centering\arraybackslash}X 
    | >{\centering\arraybackslash}X | }
    \hline
    Sample name & Number of QWs & Periods of Top DBR & Rabi Splitting (meV) \\
    \hline
    Sample P1 & 12 & 23 & $12.41 \pm 0.08$ \\  
    \hline
    Sample P2 & 12 & 29 &  $12.57 \pm 0.14$ \\  
    \hline
    Sample P3 & 12 & 32 & $13.21 \pm 0.54$  \\  
    \hline
    Sample W1 & 28 & 32 &  $13.33 \pm 0.12$ \\  
    \hline
    \end{tabularx}
    \caption{Sample designs with different quantum well (QW) numbers, top DBR periods, and Rabi splitting values derived from fits to the dispersion data near resonance. All samples have the same number of periods of the bottom DBR. Samples labeled P were grown in the Pfeiffer lab at Princeton, while the sample labeled W was grown at the Wasilewski lab at Waterloo.}  
    \label{tab:sample_designs}  
\end{table}

\section{EXPERIMENTAL METHODS}

In the experiments reported here, we used a GaAs/AlGaAs microcavity structure very similar to those of previous experiments \cite{alnatah2024coherence,alnatah2024critical,alnatah2024bose}. We have used four different samples throughout this work. Table~\ref{tab:sample_designs} summarizes their properties. Three of the four samples consisted of a total of 12 GaAs quantum wells with AlAs barriers embedded within a distributed Bragg reflector (DBR), with each sample having a different number of periods of the top DBR. The fourth sample, in contrast, consisted of a total of 28 GaAs quantum wells. In all samples, the quantum wells are in groups of four, with each group placed at one of the antinodes of cavity. Further details about the samples are discussed in the Supplementary Materials.
\par
\begin{figure*}
\includegraphics[width=0.8\textwidth]{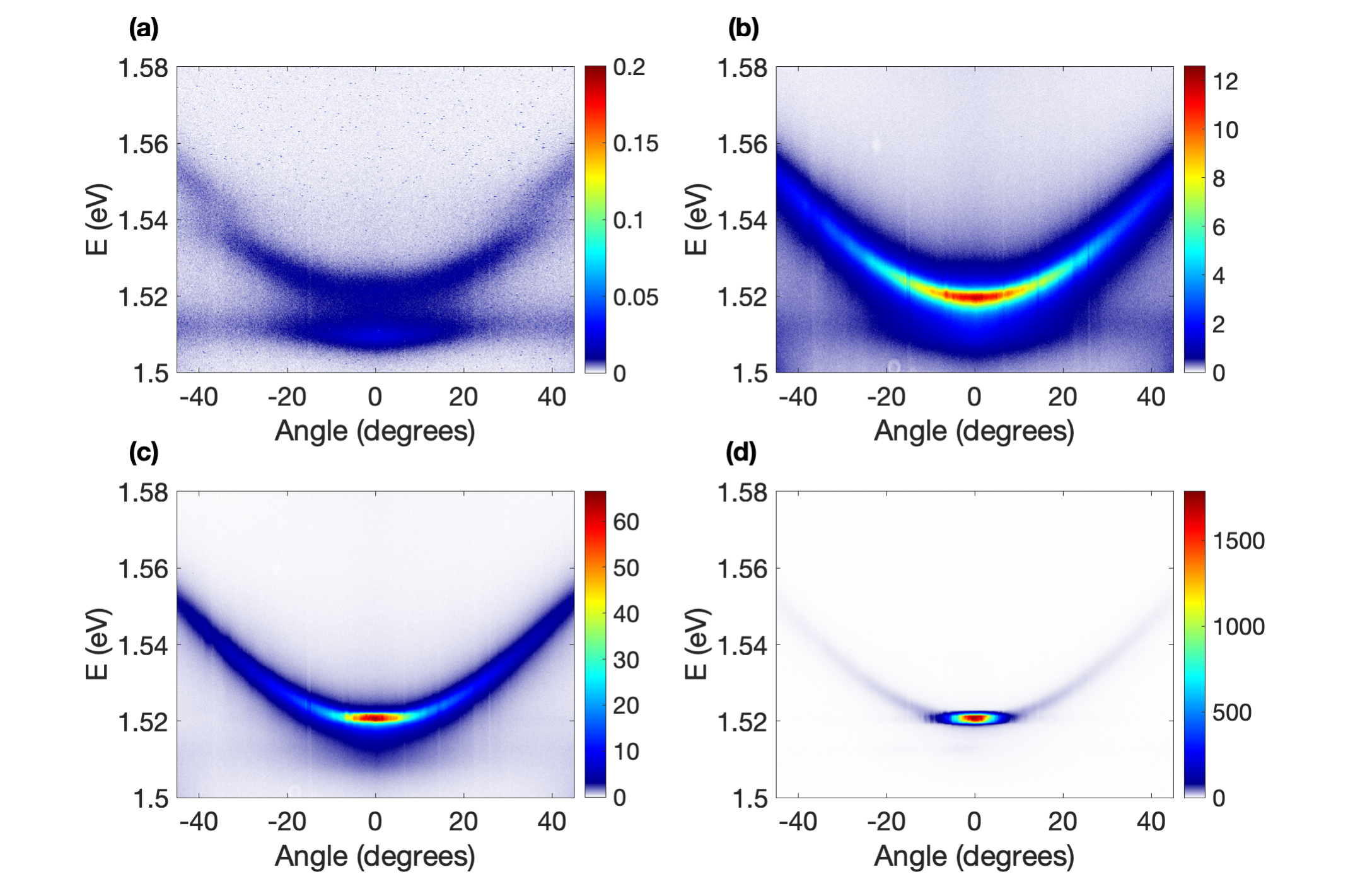}
\centering
\caption{\textbf{Angle-resolved PL for different pump powers}, at a location on Sample P1 with 0.54 exciton fraction of the LP at $k=0$. \textbf{(a)} Pump power $P=0.01P{_{\mathrm{th}}}$, \textbf{(b)}  $P = 0.83P{_{\mathrm{th}}}$, \textbf{(c)}  $P=1.00P{_{\mathrm{th}}}$ and \textbf{(d)}  $P= 1.13P{_{\mathrm{th}}}$. The threshold power $P{_{\mathrm{th}}}$ is defined in Section III of the Supplementary Material.}
\label{fig:power-series-EvsK}
\end{figure*}

All the measurements reported in this paper and supplemental material were at room temperature, with no cryogenic system. The polaritons were generated by pumping the sample non-resonantly with a wavelength-tunable laser, tuned to a reflectivity minimum approximately 172 meV above the lower polariton resonance (737.4 nm). To minimize heating of the sample, the pump laser was modulated using an optical chopper with a duty cycle of 1$\%$ and pulses of duration approximately 25 $\mathrm{\mu s}$, which is very long compared to the dynamics of the system. The non-resonant pump created electrons and holes, which scattered down in energy to become polaritons.  The photoluminescence (PL) was collected using a microscope objective with a numerical aperture of 0.75 and was imaged onto the entrance slit of a spectrometer. The image was then sent through the spectrometer to a CCD camera for time-integrated imaging. 
\par
To measure the energy dispersion of the polaritons, we used angle-resolved photoluminscence (PL) to obtain the intensity image $I(\theta,E)$, where $\theta$ is the angle of emission, which has a one-to-one mapping to the in-plane momentum of the polaritons. The angle of photon emission maps directly to the in-plane $k$-vector of the particles inside the structure. The exciton fraction can be chosen by moving to different locations on the sample, where the cavity photon energy varies because of the thickness gradient across the wafer. This thickness variation causes a change in cavity photon energy, as the wavelength of the photon inside the cavity is proportional to the cavity length.

\section{EXPERIMENTAL RESULTS}

Figure \ref{fig:strong-coupling} shows examples of dispersion of the polaritons measured at room temperature for two different samples, clearly demonstrating strong coupling with two distinct polariton branches. These branches, lower-polariton (LP) and middle-polariton (MP), arise due to the interaction between the heavy-hole exciton with the cavity photon mode. 
\par
Polariton condensation was observed as we increased the pump power across the phase transition. Figure \ref{fig:power-series-EvsK} shows angle-resolved PL for different pump powers for the case of the LP mode near resonance with the cavity photon. In this case, the higher, light-hole exciton branch, is clearly visible. As is typical for polariton condensation, the PL narrows both in $k$-space and in energy width as the density increases; the latter indicates coherence of the emission. We note that at room temperature, the two lowest polariton modes are roughly equally occupied, and therefore the lowest energy state is not strongly favored.

\par
\begin{figure}
\includegraphics[width=0.8\columnwidth]{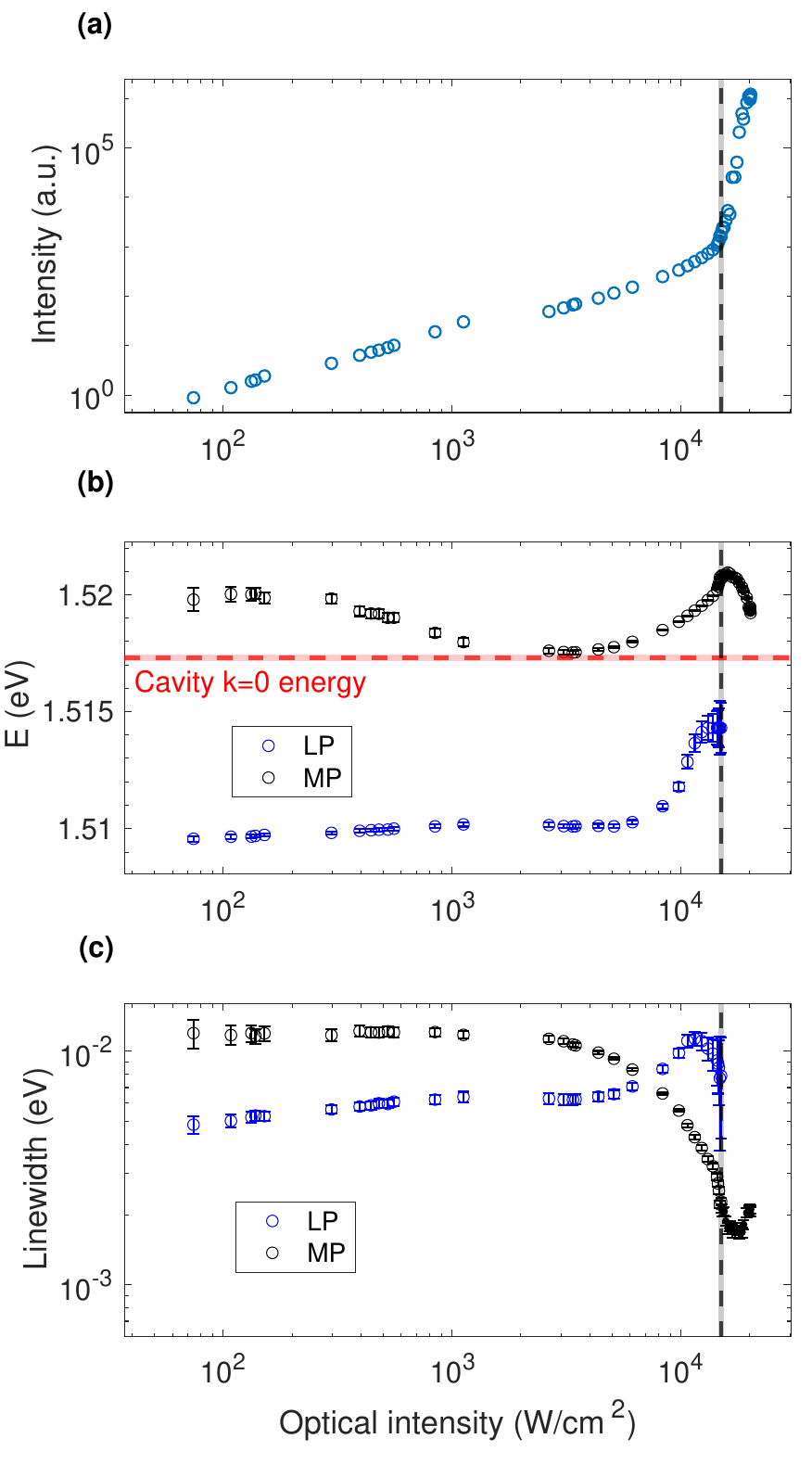}
\centering
\caption{\textbf{Blue shift and linewidth of the PL lines} for Sample P1 at a location with 0.54 LP exciton fraction at $k=0$. \textbf{(a)} The intensity at $k=0$ of the polaritons as a function of optical pump intensity, corrected for the reflection from the top surface,
\textbf{(b)} The energies of the polariton lines at $k=0$ as a function of the pump power. The red dashed line represents the cavity zero energy extracted from the three-level model fits. The line marked ``Cavity $k=0$ energy'' is the value from the fit of Figure \ref{fig:different_detunings}(b); the vertical pink range gives the uncertainty of this value. The bare heavy-hole exciton energy is very near to this value. \textbf{(c)} Circles: Full width at half maximum at $k=0$. A single Lorentzian fit was used for the data corresponding to solid circles since we see only main peaks at high power due to condensation. The vertical black dashed lines (with uncertainty given by the gray regions) indicate the power beyond which the system can no longer be identified as being in the strong-coupling regime.}
\label{fig:linewidth-blueshift}
\end{figure}

The linewidths and the energy positions of the polariton lines were extracted from the $I(\theta, E)$ images by taking a vertical slice at $\theta = 0$ to obtain $I(E)$. The resulting intensity profiles were fitted to the sum of two Lorentzians, corresponding to the LP and MP. The UP, being very excitonic at $\theta = 0$, was essentially undetectable except at the anticrossing regions, as seen in Fig.~\ref{fig:linewidth-blueshift}. Since the detuning in Fig.~\ref{fig:power-series-EvsK} corresponds to the case of approximately $0.5$ exciton fraction of the LP and MP, the difference between the peak positions of the two Lorentzians extracted from the fits, plotted in Fig.~\ref{fig:linewidth-blueshift}(c), is close to the Rabi splitting energy. Because this is a three-level system with contribution from the light-hole exciton state, the cavity photon does not lie exactly in the middle of the LP and MP states, and the splitting there is not exactly equal to the Rabi energy. 
\par
As the pump power is increased, the LP and MP lines shift but remain resolvable up to the onset of strong coherence, indicated by line narrowing and a jump in the emission intensity. At high power, only a single peak is resolvable. Above this power, we used a single Lorentzian fit to extract the linewidth and energy shift (denoted by the filled black circles).  Above the density indicated by the vertical dashed lines in Figure~\ref{fig:linewidth-blueshift}, the system appears to collapse to weak coupling, i.e., photonic condensation.
\par
The Rabi splitting at very low pumping power, extracted from the three-level model described below (with parameters provided in the Supplementary Material), ranges from 12.3 to 15.2 meV, depending on the detuning. This value is significantly larger than the heavy-hole exciton half-width, measured to be approximately 5.2 meV, yielding a ratio of $\Omega_{\mathrm{hh}}/\Gamma_{\mathrm{hh}} \approx 2.3 - 2.9$—about a factor of two above highest value reported in previous work, where $\Omega_{\mathrm{hh}}/\Gamma_{\mathrm{hh}} \approx 1-1.2$ \cite{tsintzos2009room,brodbeck2013room,suchomel2017room}. This improvement in achieving a larger $\Omega_{\mathrm{hh}}/\Gamma_{\mathrm{hh}}$ is not merely a technical enhancement, but also enables the system to maintain strong coupling at higher pumping powers. As a result, strong coupling persists well into the nonlinear regime (as shown in Fig.~\ref{fig:linewidth-blueshift}), where significant blue shift of the lower polariton line is observed, before eventually becoming weak coupling at higher pumping powers.
\par
As shown in Fig.~\ref{fig:linewidth-blueshift}, the coherence is accompanied by a non-monotonic energy shift, indicating substantial interactions present in the system and, therefore, strong nonlinearity. Above the threshold density for weak coupling, the blue shift is reversed, as the polaritons appear to be pulled back down to the bare cavity photon energy. In this same regime, the line broadens again, indicating that the system is becoming less coherent. 
\begin{figure}
\includegraphics[width=0.75\columnwidth]{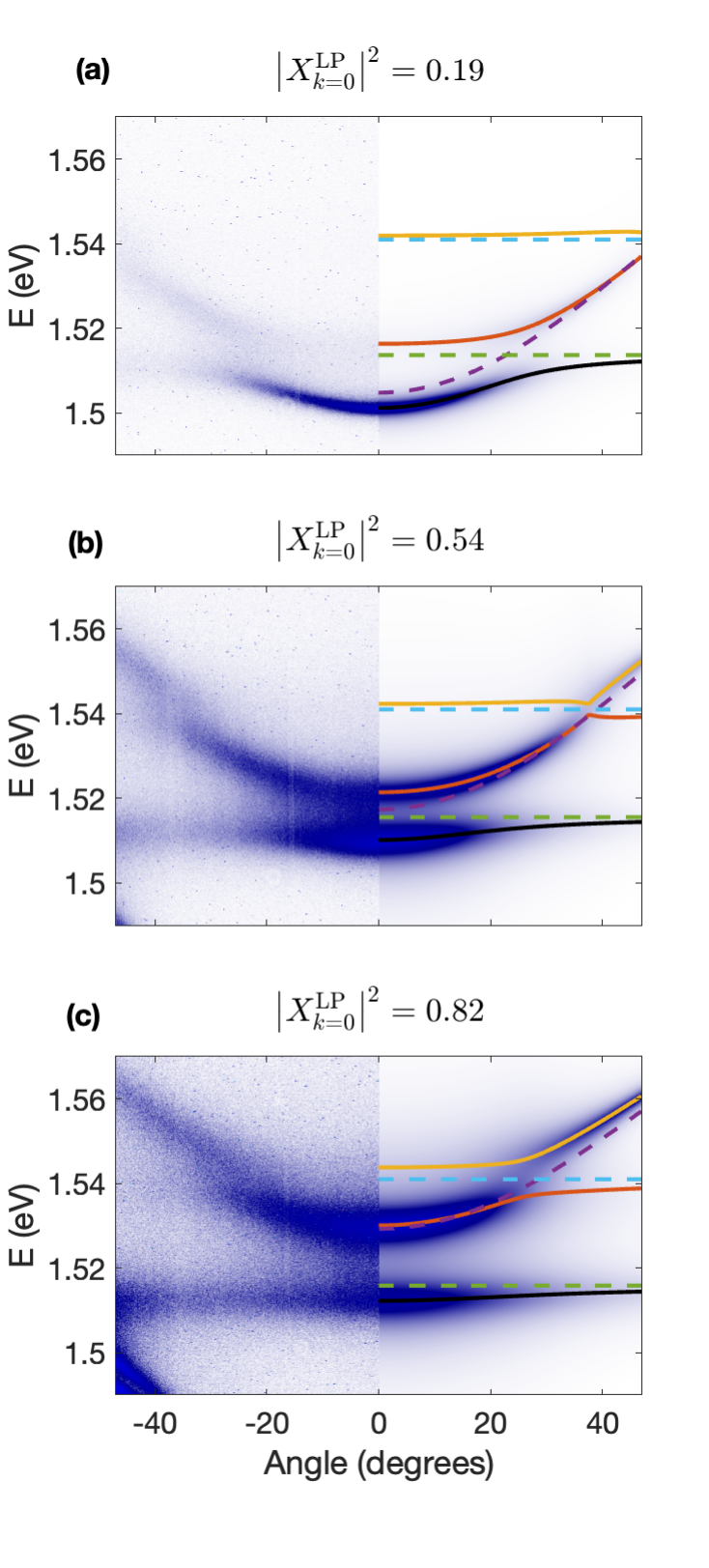}
\centering
\caption{\textbf{Angle-resolved PL for different locations on Sample P1 compared to simulation.}  Left side of each image: Angle-resolved PL measurements of the polariton at very low pumping power, corresponding to the estimated exciton fractions of the lower polariton at $k=0$ given by the upper labels. 
Right sides: simulated data using the model discussed in the text. The parameters for the fits are given in the Supplementary Material. The solid black, red, and yellow lines represent the LP, MP, and UP, respectively. The green dashed line represents the heavy-hole exciton energy, the blue dashed line represents the light-hole exciton energy, and the purple dashed line represents the cavity energy. The extra curved lines seen at low energy and high angle in (c) are ripples in the reflectivity spectrum at the edge of the stopband of the DBR cavity of this sample.}
\label{fig:different_detunings}
\end{figure}

\section{ANALYSIS}
A simple three-level model for our GaAs-based microcavity structures incorporating photons, heavy-hole excitons, light-hole excitons can be expressed as follows
\begin{equation}
H(\theta) = 
\begin{pmatrix}
 E_{\mathrm{cav}}(\theta) + i\Gamma_\mathrm{cav}& \Omega /2 & \Omega /2\\ 
\Omega /2 & E_\mathrm{hh} + i\Gamma_\mathrm{hh} & 0\\ 
\Omega /2 & 0 & E_\mathrm{lh} + i\Gamma_{\mathrm{lh}}
\end{pmatrix},
\end{equation}
where $\Gamma_\mathrm{cav}$ gives half width at half maximum of the photon linewidth broadening, which is negligible compared to the exciton linewidths in our samples,  and  
$\Gamma_{\mathrm{hh}}$ and $\Gamma_{\mathrm{lh}}$ give the heavy-hole and light-hole exciton broadening. We diaonalized the Hamiltonian for each $\theta$ value to obtain the real and imaginary parts of its eigenvalues. The imaginary parts correspond to the half widths of the polaritons. This then allowed us to compute a simulated $I(\theta,E)$ image by summing three Lorentzians, each representing the LP, MP and UP. The real parts of Hamiltonian eigenvalues give the Lorentzian peaks, while the imaginary parts give their half-widths. We ensured that the integral of each Lorentzian at each $\theta$ slice is equal to the Maxwell-Boltzmann distribution $e^{-(E(\theta)-E_{LP}(\theta =0))/k_B T}$. Finally, to account for the radiative lifetime of the polaritons, we scaled each Lorentzian by the photon fraction, since the more photonic the polariton is, the more likely it is to emit a photon outside the cavity.
\par
Figure \ref{fig:different_detunings} shows the simulations of the three-level model alongside the experimental data for different locations on the sample. The simulations successfully capture the key features observed in the experimental data, giving a similar dispersion and linewidth of the polaritons. This strong agreement highlights the effectiveness of the three-level model in capturing the essential physics governing the system. 
For photonic detunings, the light hole exciton intersects the cavity mode at a larger angle than what we can collect with our numerical aperture, making the upper polariton not visible. 

As discussed in the Supplementary Material, the transition from strong coupling to weak coupling seen at higher densities can be understood in terms of a classical dielectric model. In both limits, strong coupling and weak coupling, the nonlinear shift of the lines can be understood as an effect of the mixing of the exciton and photon states.

\par

\section{CONCLUSIONS}

The clear observation of strong coupling in a GaAs/AlGaAs structure is surprising, because the binding energy of excitons in quantum wells in this material system is only around 10 meV \cite{bastard1982exciton}, well below the thermal energy $k_BT = 25.7$~meV. This implies that there will be a significant population of free electrons and holes, which can lead to complicated renormalization of the energies of the states, but as seen clearly in these experiments, the free electrons and holes do not screen out the exciton resonance at low densities. 

In the case of resonant detuning shown here, the system clearly remains in strong coupling even as the PL line narrows into the coherent regime, until a density threshold is exceeded at which it appears to collapse to weak coupling. At detunings far from resonance, the system collapses to weak coupling at densities below the onset of coherence, but even in those cases, the exciton/electron-hole optical resonance still gives a strong interaction of the photons with excitons. This nonlinear interaction gives non-monotonic energy shifts of the emission energy in all cases. This may be useful for nonlinear optical switching applications at room temperature.

These results imply that there is potential to make exciton-polariton-based GaAs/AlGaAs devices for room-temperature applications, including low-threshold lasers and nonlinear optical devices.
\section{Acknowledgements}
This project has been supported by the National Science Foundation grant  DMR-2306977. 
\bibliography{references.bib}
\clearpage
\date{\today}
\maketitle
\beginsupplement
\title{Supplementary Materials for: Strong coupling of polaritons at room temperature in a GaAs/AlGaAs structure}
\section{SAMPLE DESIGN}
As discussed in the main text, we have used four different samples throughout this work, summarized in Table~I. Three of the four samples consisted of a total of 12 GaAs quantum wells with AlAs barriers embedded within a distributed Bragg reflector (DBR), with each sample having a different number of periods of the top DBR. The fourth sample, in contrast, consisted of a total of 28 GaAs quantum wells. In all samples, the quantum wells are in groups of four, with each group placed at one of the antinodes of cavity. Additionally, the DBRs in all four samples consisted of alternating layers of AlAs and Al$_{0.2}$Ga$_{0.8}$As.
\par
\begin{figure}
\includegraphics[width=1\columnwidth]{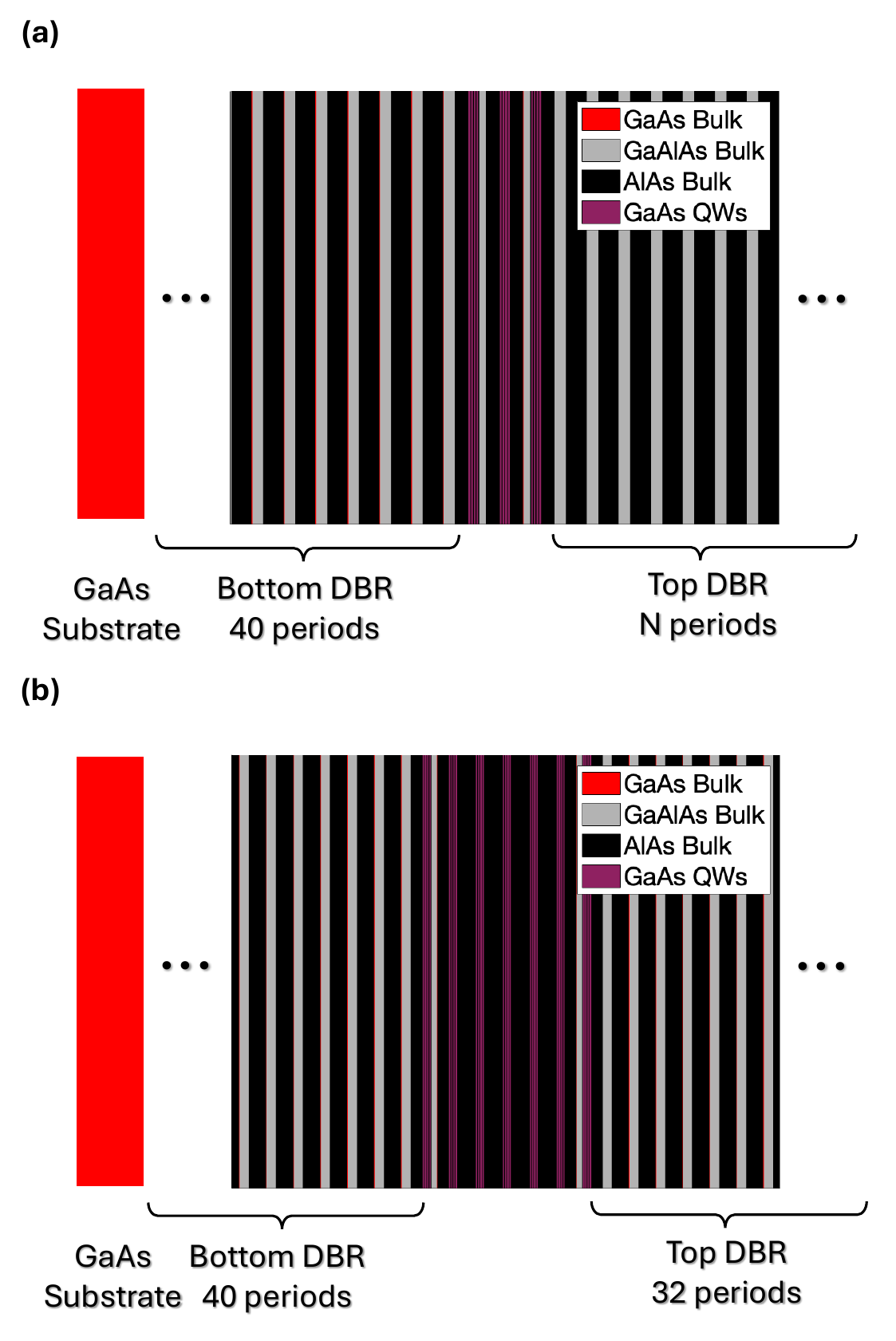}
\centering
\caption{\textbf{Sample Structures.} (a) The design of Samples P1, P2, and P3, which only differ in the number of top DBR periods $N$, with $N= 23$, $N=29$ and $N=32$, respectively. (b) The sample design of Sample W1. The QWs are arranged in groups of four, with Samples P1, P2, and P3 each containing three groups of QWs, while Sample W2 contains five groups of QWs.}
\label{fig:sample-design}
\end{figure}

Figure \ref{fig:sample-design} shows the design of Samples P1, P2, and P3, which only differ in the number of periods of the top DBR and the design of sample W1, which consisted of 28 GaAs quantum wells. 
\section{THREE-LEVEL MODEL FIT PARAMETERS}
\label{3level}
Table I summarizes the parameters obtained from the fits shown in Fig. 4 of the main text using the three-level model. In total, we extract five parameters: the energy of the heavy-hole exciton, the cavity zero energy and its refractive index, the Rabi splitting, and the temperature. In these fits, we assume that the exciton energies are independent of momentum, as excitons are much heavier than the cavity photon. For the cavity photon, we use the dispersion relation. The half-widths of the heavy- and light-hole excitons were fixed to the values measured in Fig. \ref{fig:exciton-half-width} at very low pumping power, as they are not expected to vary across different regions of the sample.
\begin{equation}
E_{\mathrm{cav}}(\theta) =  \frac{N \pi \hbar c}{L} \frac{1}{\sqrt{1-\left(\frac{\sin \theta}{n_{\mathrm{cav}}}\right)^2}} = \frac{E_{\mathrm{cav}}(\theta =0)}{\sqrt{1-\left(\frac{\sin \theta}{n_{\mathrm{cav}}}\right)^2}},
\end{equation}
where $N$ is an integer representing the number of antinodes of the cavity mode, $c$ is the speed of light and $L$ is the length of the cavity. The angle $\theta$ represents the external emission angle of the polariton, while $E_{\mathrm{cav}}(\theta =0)$ corresponds to the cavity zero energy. Lastly, $n_{\mathrm{cav}}$ is the refractive index of the cavity, which in this case is an average of the GaAs and AlGaAs materials used in the structure. 
\begin{table}[h]
    \centering
    \renewcommand{\arraystretch}{1.2}
    \resizebox{\columnwidth}{!}{%
    \begin{tabular}{lccc}
        \toprule
        Parameter & \multicolumn{3}{c}{Values} \\
        \cmidrule(lr){2-4}
                 & Fig. 4(a)  & Fig. 4(b) & Fig. 4(c) \\
        \midrule
        $E_{\mathrm{hh}}$ (eV) & 1.5137 & 1.5155 & 1.5158  \\
        $E_{\mathrm{lh}}$ (eV) & 1.5410 & 1.5410 & 1.5410 \\
        $\Gamma_{\mathrm{hh}}$ (meV) & 5.2 & 5.2 & 5.2 \\
        $\Gamma_{\mathrm{lh}}$ (meV) & 12.0 & 12.0 & 12.0 \\
        $E_{\mathrm{cav}}(\theta=0)$ (eV) & 1.5048 & 1.5173 &  1.5293 \\
        $n_{\mathrm{cav}}$ & 3.5804 & 3.6070 & 3.8899 \\
        $\Omega$ (meV) & 12.28 & 12.41 &  15.19 \\
        $T$ (K) & 159.21 & 178.49  & 200.22 \\
        \bottomrule
    \end{tabular}%
    }
    \caption{Table of parameters. The cavity photon half width $\Gamma_p$ is negligible compared to the exciton linewidths in our samples and therefore can be set to zero.}
    \label{tab:parameters}
\end{table}

\section{DEFINING CRITICAL POWER}

\begin{figure}
\includegraphics[width=0.8\columnwidth]
{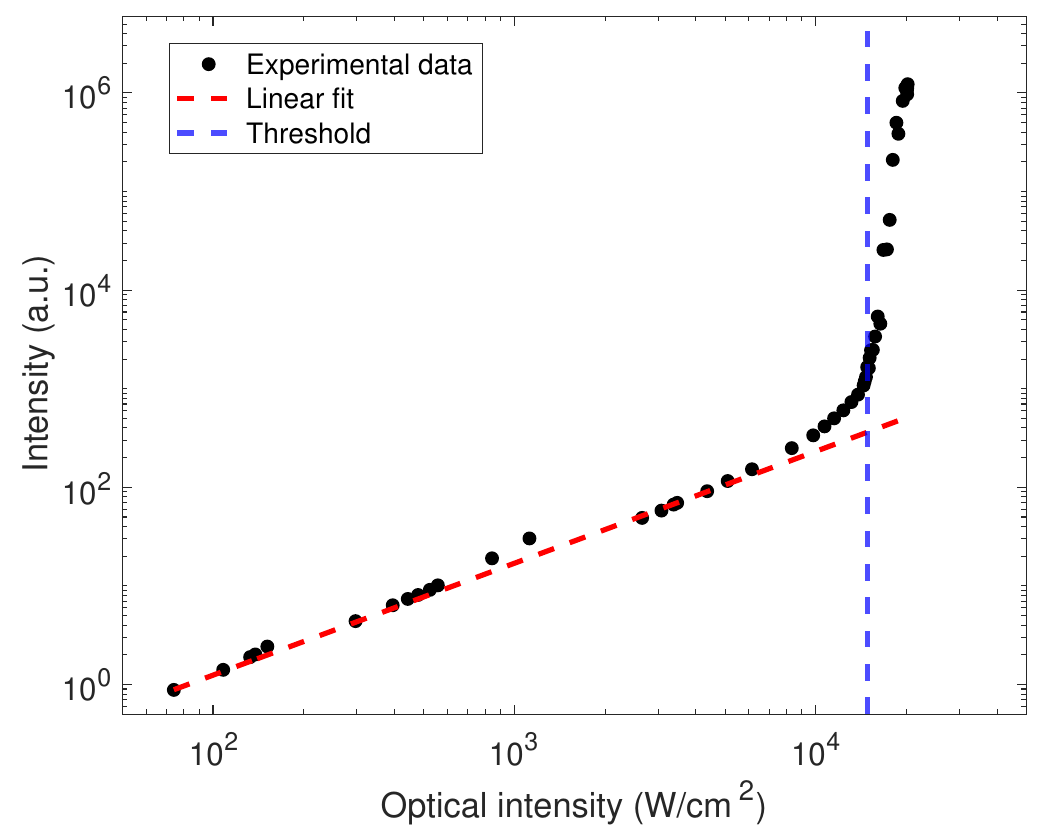}
\centering
\caption{\textbf{Condensation threshold.} The emission intensity at $k=0$ as a function of optical intensity. The blue vertical line indicates the threshold, which is defined when the the measured curve deviates from being linear by approximately $20\%$. The threshold optical intensity is approximately $1.48 \times 10^4$ \text{W/cm}$^2$.}
\label{fig:scurve}
\end{figure}

To determine threshold power of condensation, we extracted the total intensity at $k=0$ from the energy-resolved $I(\theta,E)$ images at different pump powers. The condensation threshold is identified from the characteristic ``S'' curve in Fig.~\ref{fig:scurve}, which is the same as the data of Figure~3(a) of the main text, where a nonlinear increase in intensity marks the onset of condensation.
\par
To define the threshold power, we first fitted the data in the linear regime with a linear function. The threshold was then set at the point where the measured curve deviates from the linear fit by approximately 20\%. Figure~\ref{fig:scurve} illustrates this fit and the resulting optical intensity threshold. From this method, we define the optical intensity threshold to be approximately $1.48 \times 10^4$ W/cm$^2$.

\section{ADDITIONAL DENSITY-DEPENDENCE DATA}
\begin{figure}
\includegraphics[width=0.75\columnwidth]{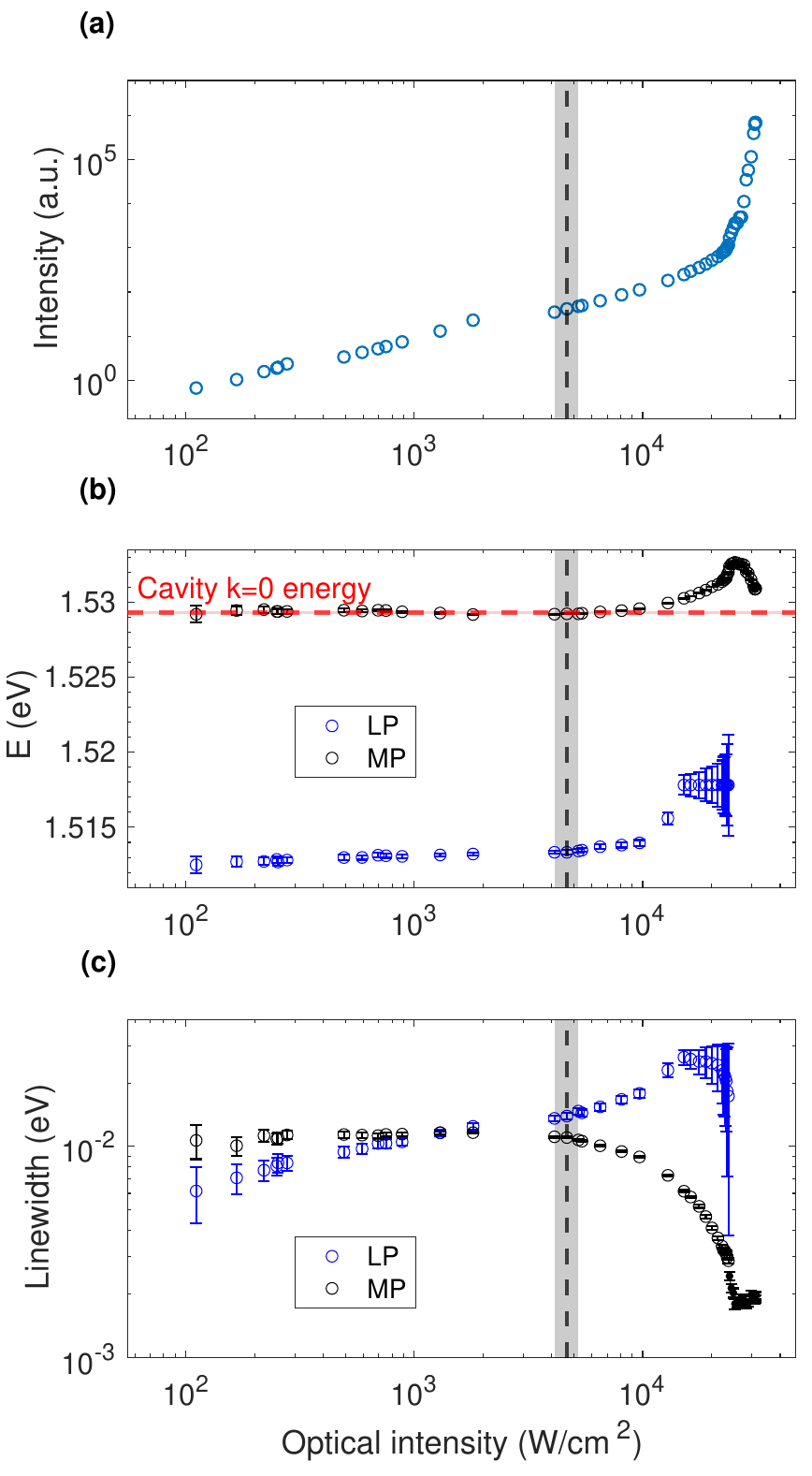}
\centering
\caption{\textbf{Blue shift and linewidth narrowing of Sample P1 for the case of 0.82 LP exciton fraction at $k=0$.} \textbf{(a)} The intensity at $k=0$ of the polaritons as a function of pump power. The threshold optical intensity as defined in Section III is 2.4 $\times 10^4$ \text{W/cm}$^2$. \textbf{(b)} The blue shift at $k=0$ as a function of the pump power. The dashed line represents the cavity zero energy extracted from the three-level model fits. The line labeled ``Cavity $k=0$ energy'' is the value from a fit like that of Figure 4(c) of the main text; the vertical pink range gives the uncertainty of this value. 
\textbf{(c)} Circles: Full width at half maximum of the lines at $k=0$. A single Lorentzian fit was used for the data corresponding to solid circles since we see only the most intense peak at high power due to condensation. The vertical black dashed line (with uncertainty given by the gray regions) indicates the power beyond which the system can no longer be identified as being in the strong-coupling regime.}
\label{fig:linewidth-blueshiftS1}
\end{figure}

\begin{figure}
\includegraphics[width=0.75\columnwidth]{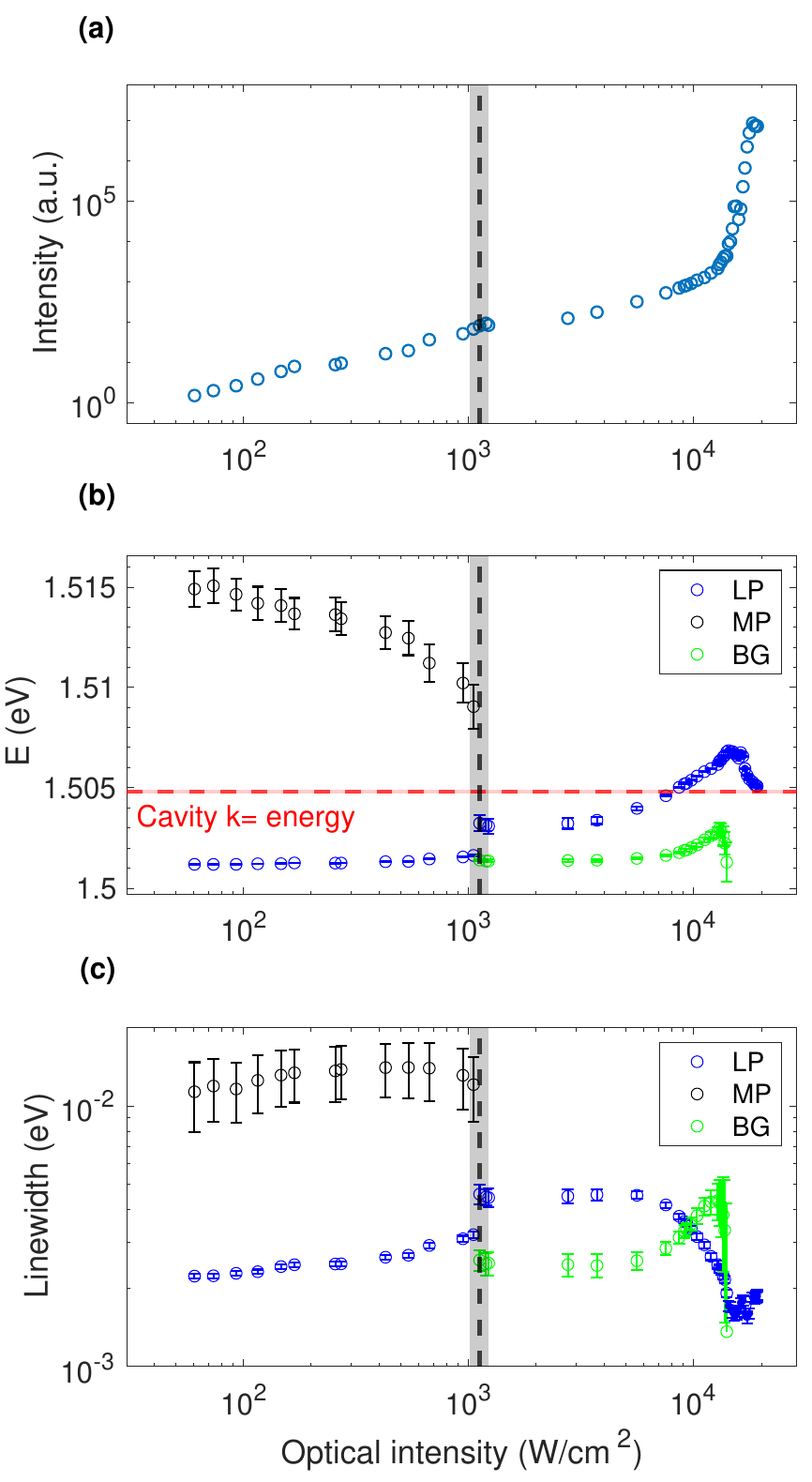}
\centering
\caption{\textbf{Blue shift and linewidth narrowing of Sample P1 for the case of 0.19 LP exciton fraction at $k=0$.}
These plots are analyzed in the same way as in Figure~\ref{fig:linewidth-blueshiftS1} and Figure 3 of the main text, but with a third, weak line below the main LP line, shown by green symbols and labeled BG for background, as discussed in the text. The threshold optical intensity as defined in Section III is 1.43 $\times 10^4$ \text{W/cm}$^2$.
 The vertical black dashed line (with uncertainty given by the gray regions) indicates the power beyond which the system can no longer be identified as being in the strong-coupling regime.} 
\label{fig:linewidth-blueshiftS2}
\end{figure}

Figures \ref{fig:linewidth-blueshiftS1} and \ref{fig:linewidth-blueshiftS2} show the same pump-power dependence curves as Figure~3 of the main text, also for Sample P1, but for two different locations with different detunings, one with the LP more photonic and one with the LP more excitonic. In the case when the LP is excitonic, the condensate occurs in the higher, photonic state, resembling a photon condensate \cite{pieczarka2024bose}. The vertical dashed lines (with uncertainty given by the gray regions) indicate where we believe the system collapse to weak coupling, primarily by observing when the anticrossing of the light-hole exciton and the photon mode at higher angle vanishes.

When the LP is mostly photonic, the case shown in Figure~\ref{fig:linewidth-blueshiftS2}, the lowest state is the brightest state. As density increases, the MP line becomes impossible to resolve as it gets closer to the LP line, and the system appears to collapse to weak coupling. At the same time, a new line with weak intensity appears below the intense LP line (labeled with green symbols). We believe this is due to the inhomogenity of the exciton region in this case. The cloud of polaritons at the periphery of the laser spot are at lower density and have a behavior similar to that of the LP states at ten times lower density. This effect, of a separate, unshifted line corresponding to a low density region around the laser spot, has often been seen previously in polariton experiments, e.g. \cite{deng2003polariton,caputo2018topological,nelsen2013dissipationless}.

In both of the cases shown here, there is a blue shift near the coherence transition even in the weak coupling limit, and then a red shift back to the bare cavity photon energy at higher density. This non-monotonic behavior is likely to be explained by renormalization of the photon energy via change of the index of refraction change. This strong renormalization, seen also at the highest densities in Figure~3 of the main text, indicates that although the system is in weak coupkling at high density, there is still a strong interaction of the photons with the excitons and free carriers.

\section{ADDITIONAL ANGLE-RESOLVED PL DATA}
\begin{figure}
\includegraphics[width=0.7\columnwidth]{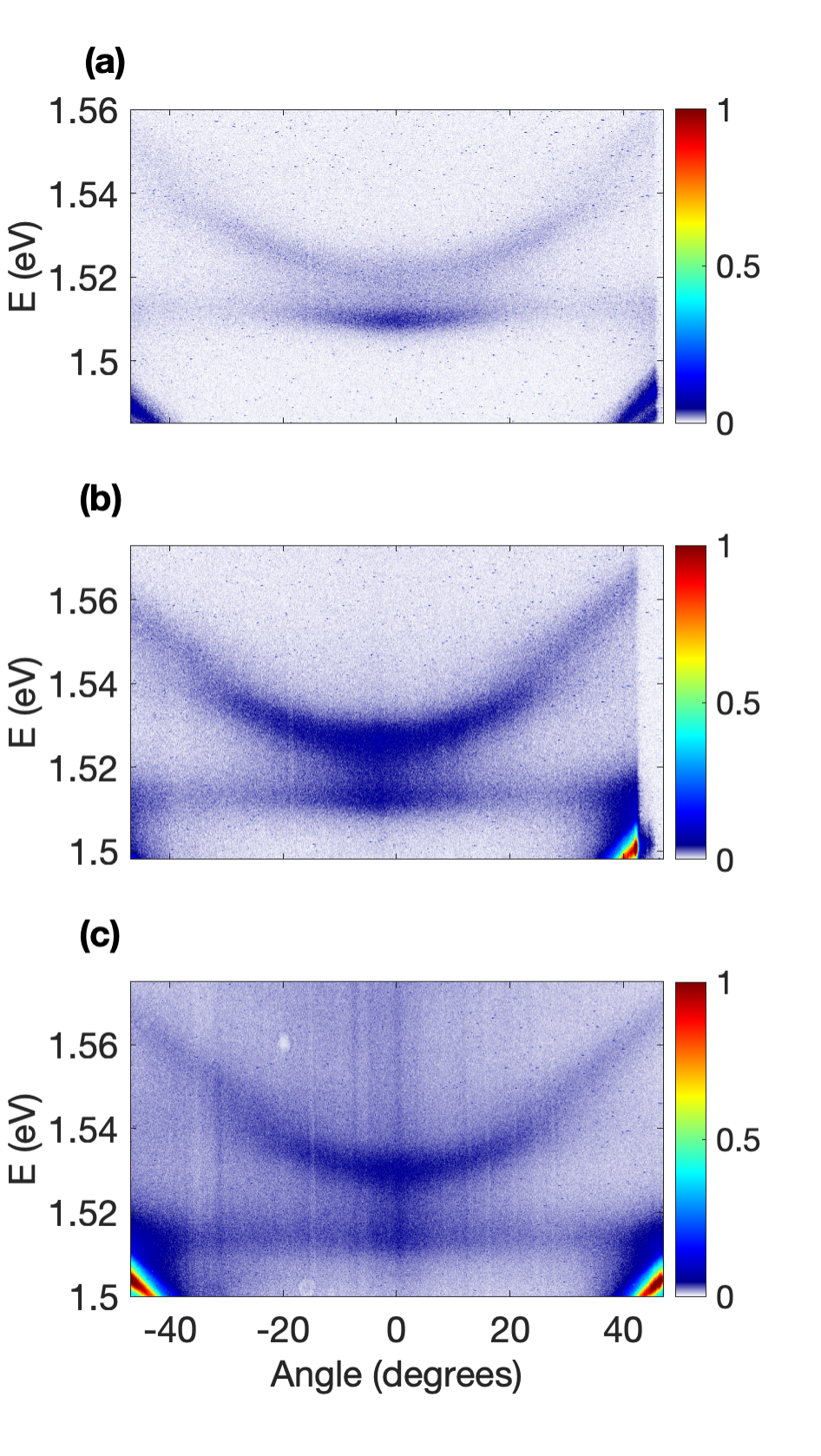}
\centering
\caption{\textbf{Angle-resolved PL for different locations on Sample P2.} The LP exciton fraction increases from (a) to (c). The extra curved lines seen at low energy and high angle are ripples in the reflectivity spectrum at the edge of the stopband of the DBR cavity of this sample.}
\label{fig:Strong-Coupling-middleQ-loren}
\end{figure}

\begin{figure}
\includegraphics[width=0.7\columnwidth]{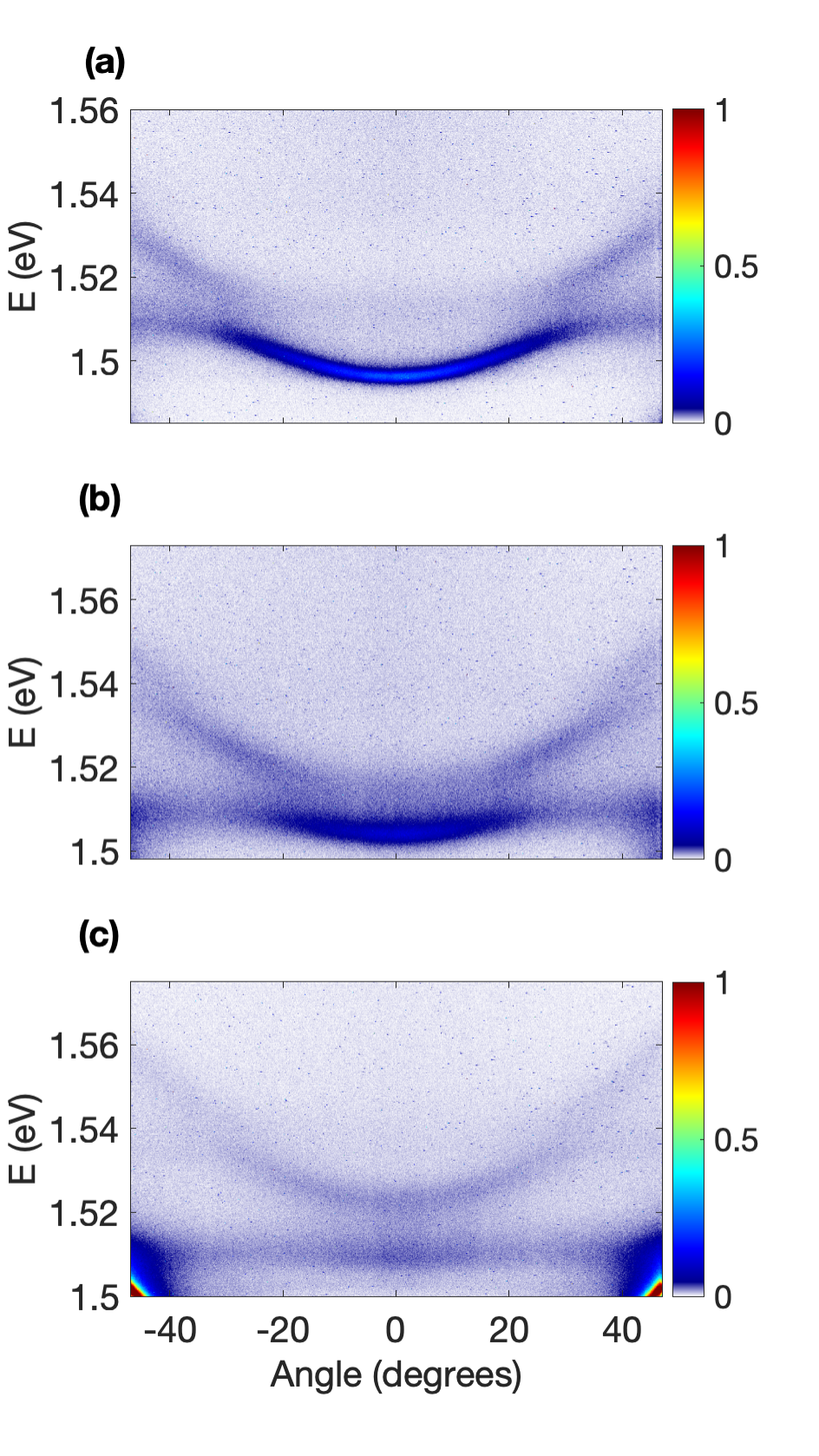}
\centering
\caption{\textbf{Angle-resolved PL for different locations on Sample P3.} The LP exciton fraction increases from (a) to (c). The extra curved lines seen at low energy and high angle are ripples in the reflectivity spectrum at the edge of the stopband of the DBR cavity of this sample.}
\label{fig:Strong-Coupling-highQ-loren}
\end{figure}

\begin{figure}
\includegraphics[width=0.7\columnwidth]{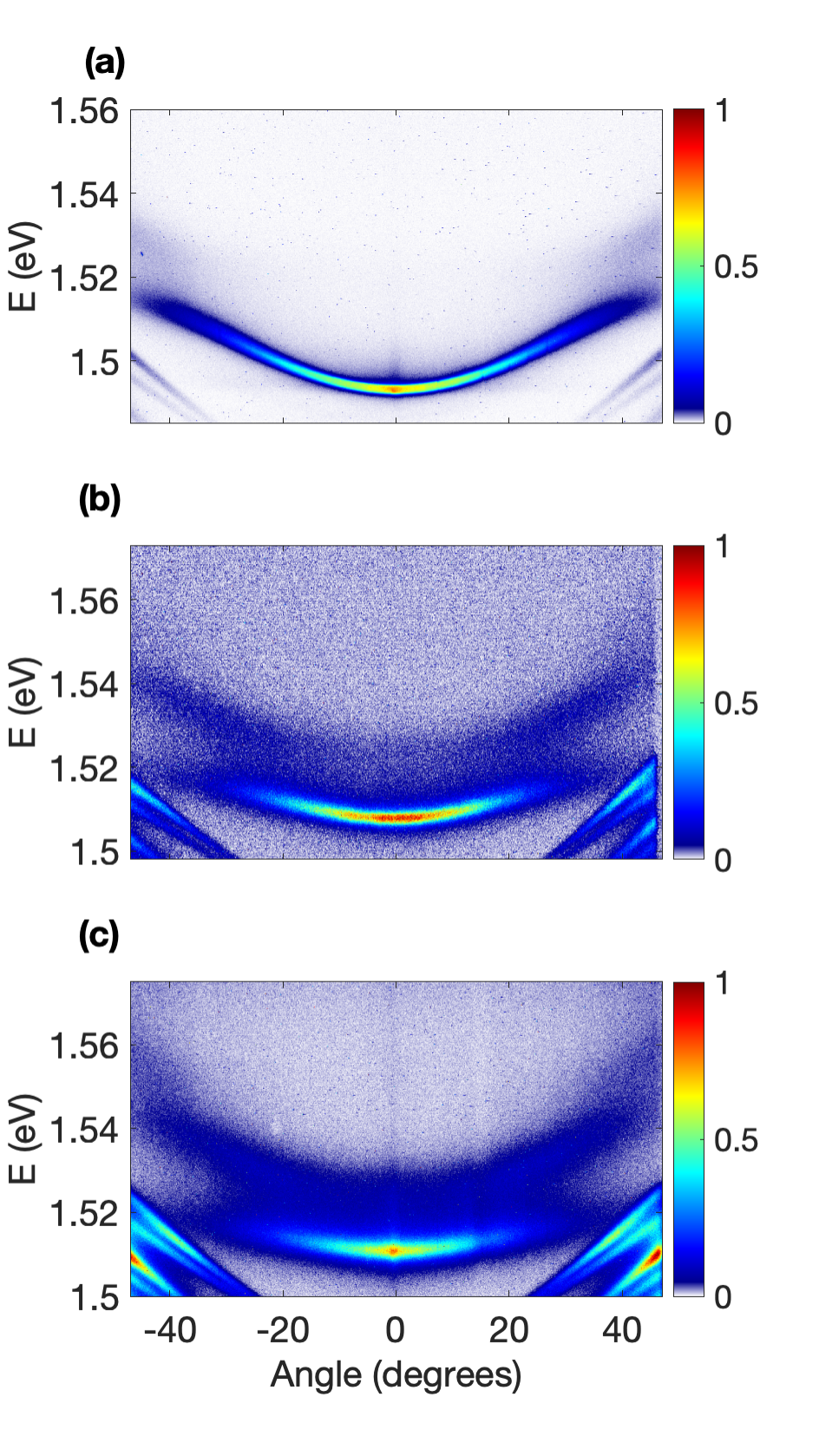}
\centering
\caption{\textbf{Angle-resolved PL for different locations on Sample W1.} The exciton fraction increases from (a) to (c). The extra curved lines seen at low energy and high angle are ripples in the reflectivity spectrum at the edge of the stopband of the DBR cavity of this sample.}
\label{fig:Strong-Coupling-Zbig}
\end{figure}

In Fig. 4 of the main paper, we reported angle-resolved images for different locations on Sample P1. Here, we report similar data for Samples P2, P3 and W1. Figures \ref{fig:Strong-Coupling-middleQ-loren}, \ref{fig:Strong-Coupling-highQ-loren} and \ref{fig:Strong-Coupling-Zbig} show the angle-resolved PL for Samples P2, P3 and W1 respectively. In each figure, we show three angle-resolved PL for different locations on the sample, ordered by increasing exciton fraction (a–c). The exciton fraction can be chosen by moving to different locations on the sample with different cavity photon energy. In all of the samples we have examined, the polaritons are in strong coupling, with well-resolved polariton branches. 


\section{Measurements of the excitons energy and linewidths}
\begin{figure}
\includegraphics[width=0.9\columnwidth]{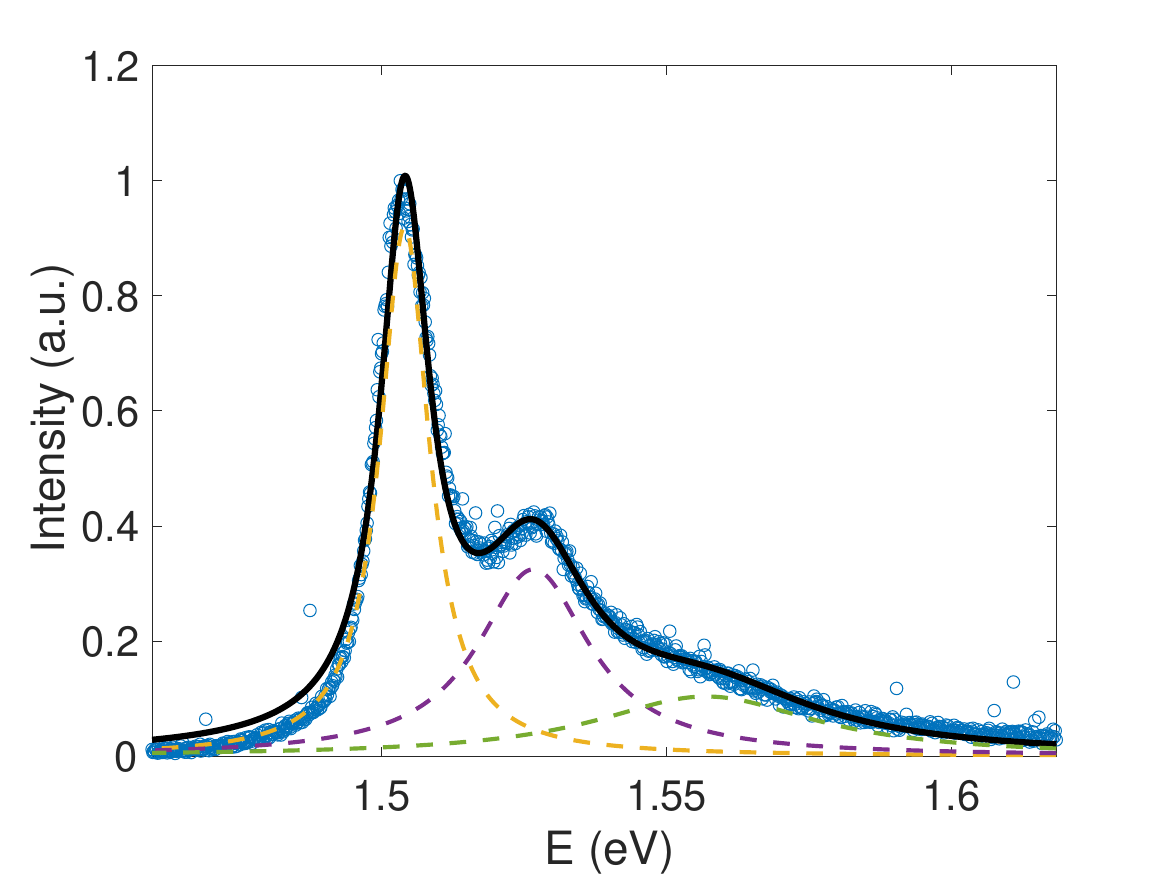}
\centering
\caption{\textbf{Excitons PL} The PL intensity $I(E)$ of excitons at very low pumping power in quantum wells at room temperature of the same design, grown directly on a GaAs substrate without DBRs. Blue circles: measured exciton PL. Black solid line: fit to the data using a sum of three Lorentzian functions. Dashed lines: individual Lorentzian functions that sum to the black solid line. The yellow dashed and purple dashed lines correspond to the heavy- and light-hole exctions respectively. The green dashed line represents the heavy hole $N=2$ quantum confined state. The fits give $\Gamma_{\mathrm{hh}} = 5.2442 $ meV and $\Gamma_{\mathrm{hh}} = 11.9656 $ meV for halfwidths of the heavy- and light-hole excitons respectively. }
\label{fig:exciton-PL}
\end{figure}

\begin{figure}
\includegraphics[width=0.9\columnwidth]{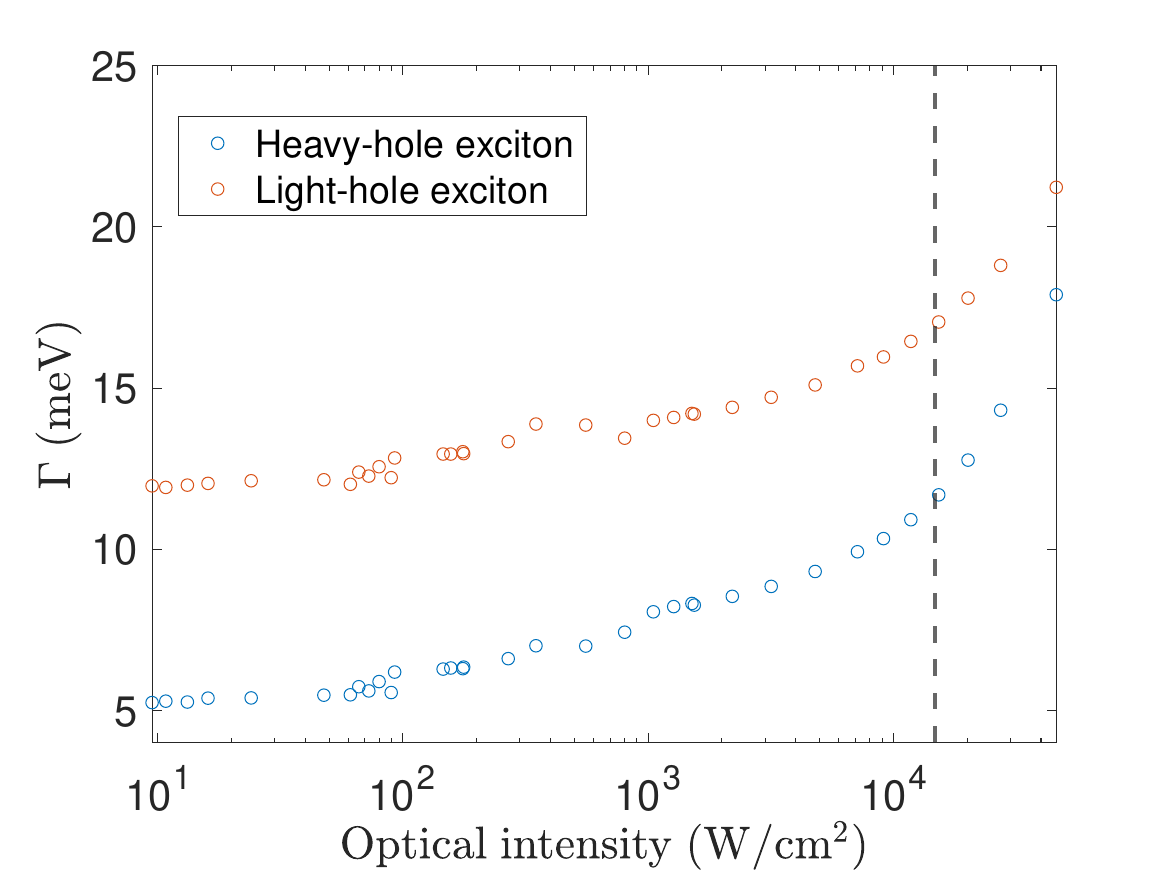}
\centering
\caption{\textbf{Exciton half-width.} The half-width of the heavy-hole and light-hole excitons in quantum wells at room temperature of the same design as a function of optical intensity of the pump, grown directly on a GaAs substrate without DBRs. The vertical line indicates the condensation threshold power near resonance, corresponding to Fig. 3 of the main text, for comparison.}
\label{fig:exciton-half-width}
\end{figure}

We have measured the PL of excitons at room temperature for a sample with bare GaAs quantum wells grown directly on a substrate grown at the Pfeiffer lab at Princeton. Figure \ref{fig:exciton-PL} shows the angle-integrated PL under non-resonant excitation, using a laser wavelength similar to that described in the main text. 
The half-widths of the excitons were extracted by fitting the PL in Fig. \ref{fig:exciton-half-width} with three Lorentzian functions. Two Lorentzian functions for the heavy- and light-hole excitons. We found that including a third Lorentzian significantly improved the fit quality. We attribute this additional peak to the $N=2$ quantum-confined state of the heavy-hole exciton. The best fit gives energies of $1.5040$ meV and $1.5569$ meV for the $N=1$ and $N=2$ heavy-hole quantum states, respectively and $1.5267$ meV for the light-hole exciton. The energy separation between these two confined states of the heavy-hole, $5.29$ meV, qualitatively agrees with the expected value for a particle in a one-dimensional infinite potential well, given by
\begin{equation}
E_{N=2}-E_{N=1} = \frac{3\hbar^2\pi^2}{2m_{\mathrm{hh}}L^2} \approx 51.16 \; \mathrm{meV},   
\end{equation}
where $L = 7 \; \mathrm{nm}$ and $m_{\mathrm{hh}} = 0.45m_e$.
\par
Figure \ref{fig:exciton-half-width} shows the extracted half-widths for the heavy and light hole excitons as a function of the pump power. As the pump power was increased, we observed line broadening of the excitons, which can be due both to exxciton-exciton interactions and to heating induced by the pump. The linewidth of the heavy-hole exciton increased by a factor of approximately 2.2, from very low pumping power to the point where polariton condensation occurs at $1-2.5\times 10^{4}\; \mathrm{W/cm^2}$ . The half-width of the heavy-hole exciton at high pumping power was approximately $12.77$ meV, which is comparable to the measured Rabi splitting discussed in the main text when the quantum wells are placed in a cavity. This is consistent with the system transitioning into the weak coupling regime. 
\par
It is important to note that the degree of light–matter coupling of polaritons can be viewed as a continuum rather than just two extremes. In the extreme weak coupling regime (i.e., $\Omega \ll \Gamma_{\mathrm{ex}}$), the eigenstates of the system are the bare photon and exciton energies. However, in the case of weak but nonzero coupling (i.e., $\Omega \sim \Gamma_{\mathrm{ex}}$), the bare photon states are not the proper eigenstates, and the eigenstates can possess a significant exciton fraction and, as a result, nonlinearity as evidenced by the significant blueshift in Fig. 3(b) of the main text.

\section{charged-oscillator model}
Although the polariton splitting is often treated as a quantum mechanical effect, as in the three-level model discussed in Section~\ref{3level}, the polariton effect can also be understood in terms of a classical dielectric constant model. (See, e.g., Ref.~\cite{snoke2023reanalysis,beaumariage2024measurement}, in which the exciton states are treated as optical dipole resonances.) It is therefore possible to model the transition from strong coupling to weak coupling using such a model. 
This allows us to see that the blue shift can be understood as a result of the renormalization of the refractive index, which is closely tied to the mixing of the exciton and photon states. 

The refractive index is renormalized because the exciton peak itself is altered by complex interactions with free electrons and phonons. We consider a simple charged-oscillator model, where the exciton susceptibility is given by
\begin{equation}
\chi_{\mathrm{ex}}(E) = A \frac{(E^2_{\mathrm{ex}}-E^2)+i\Gamma_{\mathrm{ex}} E}{(E^2_{\mathrm{ex}}-E^2)^2+\Gamma_{\mathrm{ex}}^2 E^2},    
\end{equation}
where $E_{\mathrm{ex}}$ is the exciton energy, $\Gamma_{\mathrm{ex}}$ is the FWHM of the exciton resonance and $A$ is an overall constant related to oscillator strength. To account for the heavy-hole and light-hole excitons, we sum their respective susceptibilities 
\begin{equation}
\chi_{\mathrm{QW}}(E) = \chi_{\mathrm{GaAs}} + \chi_{\mathrm{hh}}(E)+ \chi_{\mathrm{lh}}(E),    
\end{equation}
where $\chi_{\mathrm{GaAs}}$ is an overall constant that accounts for the susceptibility
of the bulk GaAs. In these simulations, we have imposed the oscillator strength of the heavy and light hole excitons to be equal $A_{\mathrm{hh}} =  A_{\mathrm{lh}}$. 
\par
\begin{figure}
\includegraphics[width=0.7\columnwidth]{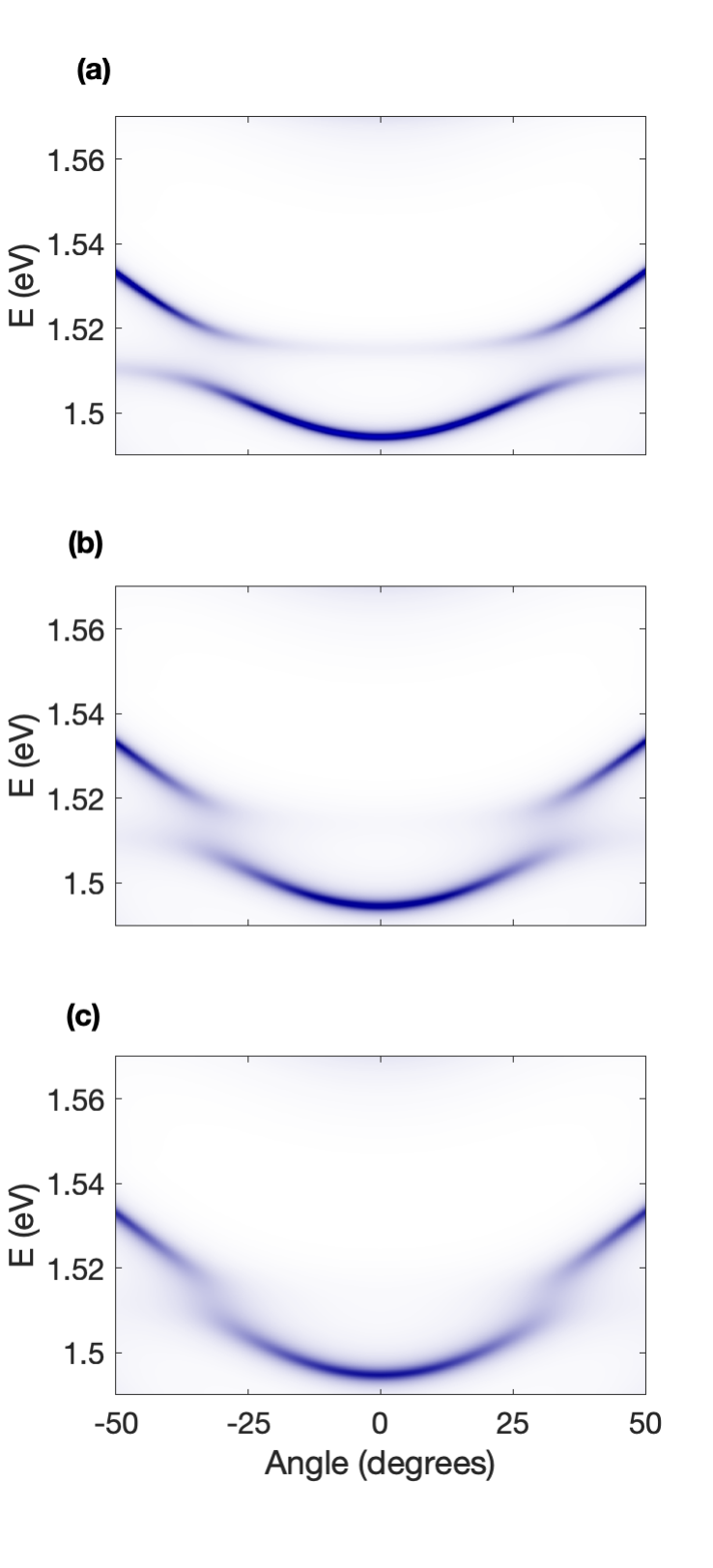}
\centering
\caption{\textbf{Transfer-matrix simulations.} The dispersion images of the polaritons using the transfer matrix method mentioned in the text. (a) Using the heavy- and light-hole exciton half-widths for the lowest pump power in Fig.~\ref{fig:exciton-half-width}, (b) using the exciton halfwidths corresponding to pump powers near the threshold of condensation given by the vertical line in Fig.~\ref{fig:exciton-half-width}, (c) using the half-widths corresponding to the highest pump power in Fig.~\ref{fig:exciton-half-width}. }
\label{fig:TMS}
\end{figure}

To generate the polariton dispersion, we follow the same procedure as in Ref. \cite{beaumariage2024measurement} by performing electromagnetic simulations using the transfer-matrix method (TMM), where the excitons are modeled as classical charged oscillators. Figure \ref{fig:TMS} shows the simulated dispersion images using the measured exciton linewidths. As expected, increasing the exciton linewidth in the simulations leads to a transition from strong to weak coupling. Notably, when using exciton linewidths measured at pump powers near the condensation threshold (indicated by the vertical dashed line in Fig. \ref{fig:exciton-half-width}), the system remains in relatively strong coupling—consistent with the experimental results shown in Fig. 3 of the main text. At much higher pump powers, corresponding to the highest pump power in Fig. \ref{fig:exciton-half-width}, the exciton linewidths broaden significantly, and the system transitions to the weak coupling regime, as illustrated in Fig.~\ref{fig:TMS}(C). 

\par
It is therefore evident that there is not a sharp distinction between the polariton model, with strong coupling, and renormalization of refractive index of the cavity due to the effect of carrier density on the exciton resonance. 

\section{THERMALIZATION DATA}

\begin{figure}
\includegraphics[width=0.9\columnwidth]{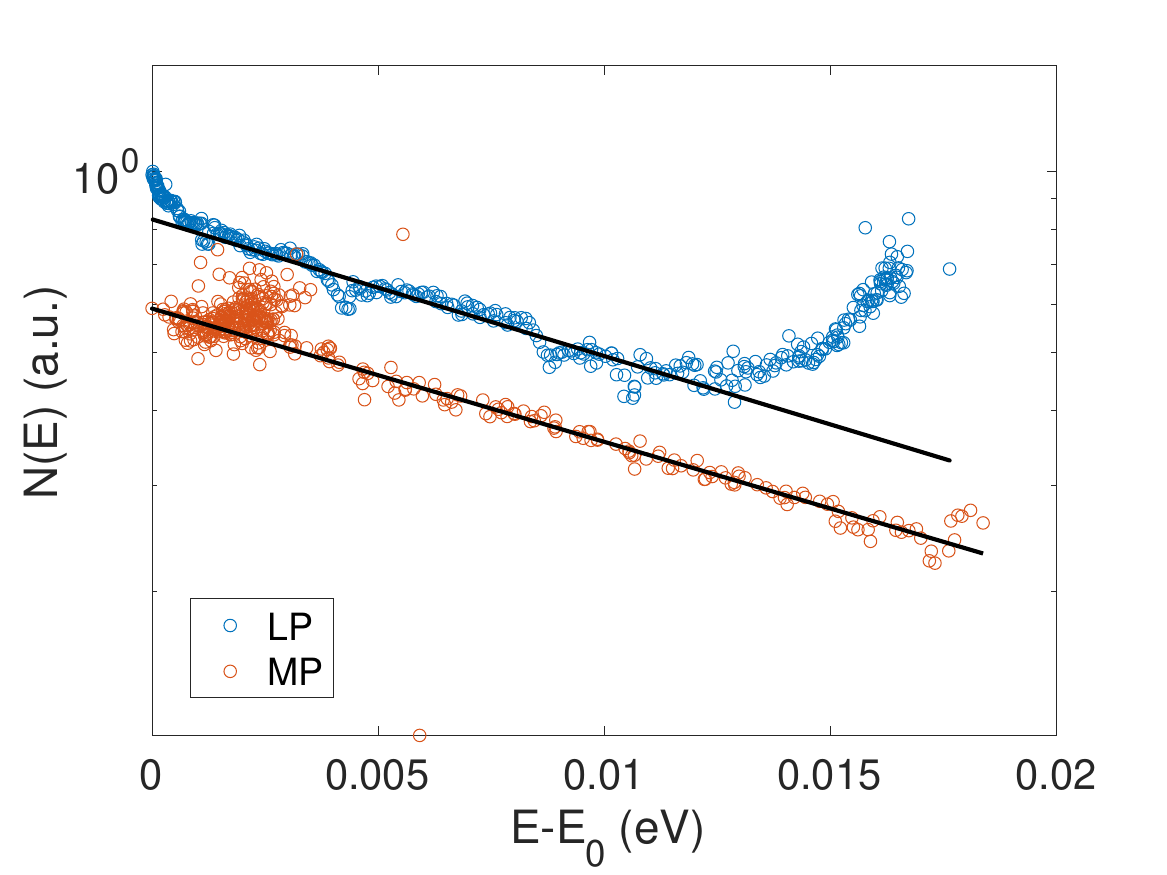}
\centering
\caption{\textbf{Occupation number.} The occupation of the LP and MP extracted from angle-resolved images for Sample P1 corresponding to the case of LP exciton fraction of $0.1$ at k=0, at low polariton density. The straight black lines correspond to Maxwell-Boltzmann distributions with temperatures of 227.3 K for the MP and 221.6 K for the LP.  }
\label{fig:thermal}
\end{figure}

Figure \ref{fig:thermal} shows the average occupation number of the MP and LP for a location on Sample P1 with photonic detuning. Other detunings have similar behavior, but the third, light-hole excitonic branch visible in those cases makes the analysis have greater uncertainty. The data was extracted from angle-resolved images like Figure~1 of the main text. At each angle, corresponding to a value of in-plane $k$-vector, a slice $I(\theta,E)$ was taken, and this was fit using a sum of Lorentzians as in Figure~1 of the main text. The spectral weight of each Lorentzian was then divided by the photon fraction for that state found from fits like those of Figure~4 of the main text, and this was assigned to the center energy of that Lorentzian. This then gave a value proportional to $N(E)$ for each state. The data scatter of the MP near $E=0$  arises because the MP is highly excitonic at $k=0$, leading to a weak peak that is difficult to fit accurately. 

As seen in this figure, the fit to an equilibrium temperature is good for both the MP and LP branches at low energy. In the LP state, at high $k$-vector, the occupation deviates strongly from the equilibrium prediction. This occurs in the ``polariton bottleneck'' region where the dispersion has an effective negative mass. 

We believe that the temperature $\sim 200$~K, below room temperature, also seen for a photon condensate \cite{pieczarka2024bose}, is due to the fact that the carriers injected by the optical pump lose energy rapidly by emitting many optical phonons with energy 36.6 meV, which can leave the polaritons with an energy below $k_BT$, after which they can only slowly heat up by acoustic phonon absorption. The cascade of optical phonon emission has been seen in earlier work \cite{kash1989carrier}, as well as cooling due to rapid optical phonon emission \cite{snoke1992evolution}.

\section{GAIN CURVES}
\begin{figure}
\includegraphics[width=0.9\columnwidth]{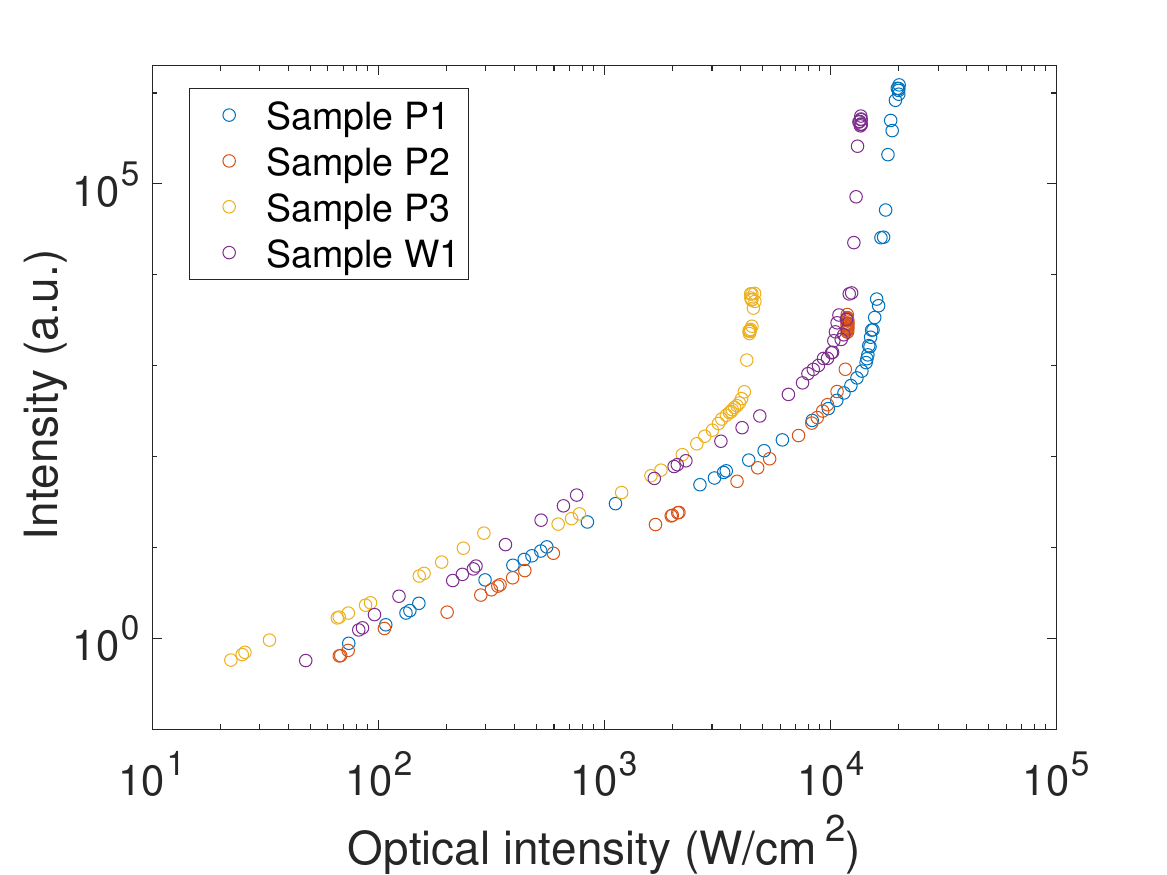}
\centering
\caption{\textbf{Gain curves.} The intensity at $k=0$ as a function of pump optical intensity extracted for
the case of the LP mode near resonance for all four samples.}
\label{fig:gain}
\end{figure}

Figure~\ref{fig:gain} shows the emission at $k =0$ as a function of pump intensity into the cavity (normalized for the surface reflection in each case), similar to Figure 3(a) of the main text and Fig.~\ref{fig:linewidth-blueshiftS1}(a) and \ref{fig:linewidth-blueshiftS2}(b) above, for all four samples, for detuning near 50\% excitonic fraction of the LP. The pump intensity was calculated as the pump power divided by the measured area of the laser excitation spot used in each case.

The effect of increasing the cavity $Q$-factor is quite clear: the intensity threshold is  lower for higher $Q$, even for comparable Rabi splitting. This is expected since the higher $Q$ corresponds to longer polariton lifetime, which boosts the density at a given pump intensity. The samples with smaller $Q$-factor have much higher threshold, but the exact value of the threshold may also depend on other structural factors.

\end{document}